\documentstyle[12pt]{article}

\def\be{\begin{eqnarray}}
\def\ee{\end{eqnarray}}
\def\bq{\begin{equation}}
\def\eq{\end{equation}}
\def\ben{\begin{enumerate}}\def\een{\end{enumerate}}
\def\d {\partial}

\def\Tr{\rm Tr}
\def\prl {Phys. Rev. Lett.}
\def\np{Nucl. Phys.}\def\pl{Phys. Lett.}
\def\la{\langle}\def\ra{\rangle}

\def\roughly#1{\mathrel{\raise.3ex\hbox{$#1$\kern-.75em%
\lower1ex\hbox{$\sim$}}}}

\begin{document}
\begin{titlepage}
\hfill

\vspace{.2cm}
\begin{center}
\ \\
{\large \bf Color Anomaly and Flavor-Singlet Axial Charge\\ of 
the Proton in the Chiral Bag:}

{\bf  The Cheshire Cat Revisited}
\ \\
\ \\
\vspace{1.0cm}
{Mannque Rho$^{(a,b)}$ and Vicente Vento$^{(a,c)}$}
\vskip 0.5cm

{\it (a) Departament de Fisica Te\`orica}

{\it Universitat de Val\`encia}

{\it 46100 Burjassot (Val\`encia), Spain}

{\it (b) Service de Physique Th\'eorique, CEA Saclay,}

{\it F-91191 Gif-sur-Yvette, France}
\vskip 0.3cm

\today

\end{center}
\vskip 0.6cm
\centerline{\bf ABSTRACT}
\vskip 0.6cm
Quantum effects inside the chiral bag induce a color anomaly which 
requires a compensating surface term to prevent breakdown of color 
gauge invariance. We show that the presence of this
surface term first discovered several years ago allows one to derive
in a gauge-invariant way a chiral-bag version of the Shore-Veneziano two-component formula for the flavor-singlet 
axial charge of the proton. This has relevance to 
what is referred to
as the ``proton spin problem" on the one hand and to the Cheshire-Cat
phenomenon in hadron structure on the other. We show that when
calculated to the
leading order in the color gauge coupling and for a specific color electric
monopole configuration in the bag, one can obtain a striking Cheshire-Cat
phenomenon with a negligibly small singlet axial charge.

\vspace{2cm}
\noindent{\small (c) Supported in part by DGICYT-PB94-0080}
\end{titlepage}

\section{Introduction}
\indent\indent
It was discovered some years ago \cite{NRWZ1,NRWZ2} that the
vacuum fluctuation inside a chiral bag that induces the
leakage of baryon charge into the hedgehog pion field outside induces color
leakage if one allows for a coupling to a 
pseudoscalar isoscalar field $\eta^\prime$. This would break color gauge 
invariance in the model unless
it is canceled by a surface counter term of the form (which will be
referred to as NRWZ counter term in what follows)
\bq
{\cal L}_{CT}=i\frac{g^2}{32\pi^2}\oint_{\Sigma} d\beta K^\mu n_\mu
({\Tr}{\rm ln} U^\dagger -{\Tr} {\rm ln} U)\label{LCT}
\eq
where $N_F$ is the number of flavors (here taken to be =2), $\beta$ is a point on a surface $\Sigma$, $n^\mu$ is the outward 
normal to the bag surface,
$U$ is the $U(N_F)$ matrix-valued field written as
$U=e^{i\pi/f} e^{i\eta/f}$ and $K^\mu$
the properly regularized Chern-Simons current
$K^\mu=\epsilon^{\mu\nu\alpha\beta} (G_\nu^a G_{\alpha\beta}^a
-\frac 23 f^{abc} g G_\nu^a G_\alpha^b G_\beta^c)$ given in terms of the
color gauge field $G^a_\mu$. Note 
that (\ref{LCT}) manifestly breaks color gauge invariance, so the action of the chiral
bag model with this term is not gauge invariant but as shown in \cite{NRWZ1}, when quantum fluctuations are calculated, there appears an induced anomaly term on the
surface which exactly cancels this term. Thus gauge invariance is restored
at the quantum level. 

In this paper, we show
that a proper account of this term allows us to formulate a fully consistent
gauge invariant treatment of the flavor-singlet axial current (FSAC) 
matrix element of the proton removing a serious
conceptual error committed in the previous work done by us
together with Park and Brown \cite{PVRB,PV}. In the work of refs.\cite{PVRB,PV},
the axial anomaly in the FSAC was introduced explicitly in terms of
a Chern-Simons current inside the bag and of a heavy $\eta^\prime$ field
(which we shall denote simply $\eta$ in the equations) outside the bag arguing 
that the diagonal matrix element was gauge invariant while off-diagonal
terms are not. However this argument strictly speaking is incorrect although
it turns out that the conclusion reached there remains more or less correct.
See Cheng \cite{cheng} for a recent discussion on this point.
In this paper, we propose to formulate the theory without invoking 
{\it ab initio} the problematic Chern-Simons current inside the bag. For this, the NRWZ color boundary condition plays a crucial role.

A complete description calls for a full Casimir calculation which is 
highly subtle and yet to be 
performed. In this paper, we shall limit ourselves
to the lowest non-trivial order
in the color gauge coupling constant and find that the Cheshire Cat principle
-- that physics should be more or less independent of the confinement
bag radius\cite{NRZ} --  found in the non-anomalous sector is also applicable 
in the anomalous sector 
for a particular field configuration for the color electric 
field. This conclusion differs from that of Dreiner, Ellis and Flores \cite{DEF}
who obtained the opposite result by
ignoring the perturbative gluon effect inside the bag. We shall see that the
Dreiner-Ellis-Flores scenario can be recovered in a particular limit of
our theory.

This paper is organized as follows. In Section 2, the formulation of the
theory implementing the color anomaly is presented. The relevant axial
charge with the $U_A (1)$ anomly suitably incorporated is computed
in terms of the degrees of freedom that figure in the chiral bag model.
In Section 3, the results of the calculation are given. Further
discussions and open problems are found in Section 4.

\section{Formulation}
\subsection{Boundary conditions}
\indent\indent
The equations of motion for the gluon and quark fields
inside and the $\eta^\prime$ field outside are the same as in
\cite{PVRB,PV}. However the boundary conditions on the surface now 
read
\bq
\hat{n}\cdot \vec{E}^a=-\frac{N_F g^2}{8\pi^2 f} \hat{n}\cdot \vec{B}^a 
\eta\label{E}
\eq
\bq
\hat{n}\times \vec{B}^a=\frac{N_F g^2}{8\pi^2 f} \hat{n}\times \vec{E}^a 
\eta\label{B}
\eq
and
\bq
\frac 12 \hat{n}\cdot({\bar{\psi}}{\bf \gamma}\gamma_5\psi)
=f \hat{n}\cdot\d \eta +C \hat{n}\cdot K\label{bc}
\eq
where $C=\frac{N_F g^2}{16\pi^2 }$ and $\vec{E}^a$ and $\vec{B}^a$ are,
respectively, the color electric and color magnetic fields. 
Here $\psi$ is the QCD quark field.

As it stands, the boundary condition for the $\eta^\prime$ field (\ref{bc})
looks gauge non-invariant because of the presence of the normal
component of the Chern-Simons current on the surface. However
this is not so. As shown in \cite{NRWZ2}, the term on the LHS of (\ref{bc})
is
not well-defined without regularization and when properly regularized,
say, by point-splitting, it can be written in terms of
a well-defined term which we will write as $\frac 12:{\bar{\psi}}\hat{n}\cdot
\gamma\gamma_5\psi:$ plus a gauge non-invariant term (see eq.(2) of 
\cite{NRWZ2}) which cancels exactly 
the second term on the RHS. The resulting boundary condition 
\bq
\frac 12\hat{n}\cdot:({\bar{\psi}}{\bf \gamma}\gamma_5\psi):
=f \hat{n}\cdot\d \eta \label{bc2}
\eq
is then perfectly well-defined and gauge-invariant. However it is useless
as it stands since there is no simple way to evaluate the left-hand
side without resorting to a model.
Our task in the chiral bag model is to 
express the well-defined operator $:(\bar{\psi}
\vec{\gamma}\gamma_5\psi):$ in terms of the bagged quark field $\Psi$. 
In doing this, our key strategy is to eliminate gauge-dependent surface terms
by the NRWZ surface counter term.

\subsection{Flavor-singlet axial current}
\indent\indent
Let us write the flavor-singlet axial current in the model as a sum of
two terms, one from the bag and the other from the outside populated by
the meson field $\eta^\prime$ (we will ignore the Goldstone pion fields for 
the moment)
\bq
A^\mu =A^\mu_B \Theta_B + A^\mu_M \Theta_M.\label{current}
\eq
We shall use the short-hand notations $\Theta_B=\theta (R-r)$ and
$\Theta_M=\theta (r-R)$ with $R$ being the radius of the bag which we shall
take to be spherical in this paper.
We demand that the $U_A (1)$ anomaly be given in this model by
\bq
\partial_\mu A^\mu = 
\frac{\alpha_s N_f}{2\pi}\sum_a \vec{E}^a \cdot \vec{B}^a \Theta_{B}+
f m_\eta^2 \eta \Theta_{M}.\label{ABJ}
\eq
Our task is to construct the FSAC in the chiral bag model
that is gauge-invariant and consistent with this anomaly equation. 
Our basic assumption is that in the nonperturbative sector outside of the
bag, the only relevant $U_A (1)$ degree of freedom is the massive $\eta^\prime$
field. (The possibility that there might
figure additional degrees of freedom in the exterior of the bag co-existing
with the $\eta^\prime$ and/or inside the bag 
co-existing with the quarks and gluons
will be discussed later.)
This assumption allows us to write
\bq
A^\mu_M = f\d^\mu \eta
\eq
with the divergence  
\be
\d_\mu A^\mu_M &=& fm_\eta^2 \eta.
\ee
Now the question is: what is the gauge-invariant and regularized
$A^\mu_B$ such that the anomaly (\ref{ABJ}) is satisfied?
To address this question, we rewrite the current (\ref{current})
absorbing the theta functions as
\bq
A^\mu=A_1^\mu +A_2^\mu
\eq
such that
\be
\partial_\mu A_1^\mu &=& f m_\eta^2 \eta \Theta_{M},\label{Dbag}\\
\partial_\mu A_2^\mu &=& 
\frac{\alpha_s N_f}{2\pi}\sum_a \vec{E}^a \cdot \vec{B}^a \Theta_{B}.
\label{Dmeson}
\ee
We shall deduce the appropriate currents in the lowest order in the 
gauge coupling constant $\alpha_s$
and in the cavity approximation for the quarks
inside the bag. 
\subsubsection{The ``quark'' current $A_1^\mu$}
\indent\indent
Let the bagged quark field be denoted $\Psi$. Then to the {\it lowest order}
in the gauge coupling and ignoring possible additional degrees of
freedom alluded above, the boundary condition (\ref{bc2}) is 
\bq
\frac 12\hat{n}\cdot({\bar{\Psi}}{\bf \gamma}\gamma_5\Psi)
=f \hat{n}\cdot\d \eta \label{bc3}
\eq
and the corresponding current satisfying (\ref{Dbag}) is
\bq
A_1^\mu=A_{1q}^\mu +A_{1\eta}^\mu
\eq
with
\be
A_{1q}^\mu &=& (\bar{\Psi} \gamma^\mu \gamma_5\Psi)\Theta_B,\label{A1in}\\
A_{1\eta}^\mu &=& f\d^\mu \eta \Theta_M.\label{A1out}
\ee
We shall now proceed to obtain the explicit form of
the bagged axial current operator.
In momentum space, the quark contribution is
\be
A^j_{1q}(q) &=& \frac{1}{2} \int d^3r e^{i \vec{q} \cdot \vec{r}}
\la N_{Bag}|\Psi^\dagger\sigma^j\Psi|N_{Bag}\ra\nonumber\\
&=& \left(a(q) \delta_{j\,k} +b(q)(3 \hat{q}_{j} \hat{q}_{k}
- \delta_{j\,k})\right) \la \frac{1}{2} \sum_{quarks} \sigma^k\ra\label{A1q}
\ee
where
\be
a(q)& =& N^2 \int dr r^2 (j_0^2(\omega r) -\frac{1}{3}j_1^2(\omega r))
j_0(q r),\label{MITga}\\
b(q) &=& \frac{2}{3} N^2 \int dr r^2 j_1^2(\omega r) j_2(q r)
\ee
where $N$ is the normalization constant of the (bagged) quark wave function.
In the limit that $q\rightarrow 0$ which is what 
we want to take for the axial charge, both terms are non-singular 
and only the $a(0)$ term survives, giving
\bq
A^j_{1q}(0) = g_{A,quark}^0  \la \frac{1}{2} \sum_{quarks} \sigma^j\ra
\label{1q}
\eq
where $g_{A,quark}^0$ is the singlet axial charge of the bagged quark
which can be extracted from (\ref{MITga}). In the numerical estimate
made below, we shall include the Casimir effects associated with
the hedgehog pion configuration to which the quarks are coupled
\cite{casimir,falomir}, so
the result will differ from the naive formula (\ref{MITga}).

To obtain the $\eta^\prime$ contribution, we take the $\eta^\prime$ 
field valid for 
a static source
\be
\eta (\vec{r}) = - \frac{g}{4\pi M} \int d^3r'\chi^{\dagger} \vec{S}
\chi \cdot \vec{\nabla} \frac{e^{-m_{\eta}|\vec{r} - \vec{r'}|}}
{|\vec{r} - \vec{r'}|}
\ee
where $g$ is the short-hand for the $\eta^\prime NN$ coupling constant,
$M$ is the nucleon mass, $\chi$ the Pauli spinor for the nucleon and 
$S$ the spin operator. The contribution to the FSAC is
\be
A^j_{1\eta}(q) &=& \int_{V_M} d^3r  e^{i \vec{q} \cdot \vec{r}}f 
\partial_j \eta,\nonumber\\
&=& (c(q) \delta_{j\,k} +d(q)(3 \hat{q}_{j} \hat{q}_{k}
- \delta_{j\,k})) \la \frac{1}{2} \sum_{quarks} \sigma^k\ra
\ee
with
\be
c(q) &=& \frac{fg}{2M}\int_R^\infty drr^2 \frac{e^{-m_\eta r}}{r} m_\eta^2 
j_0 (qr),\label{cq}\\
d(q) &=& -\frac{fg}{2M}\int_R^\infty dr \frac{e^{-m_\eta r}}{r}
[r^2 m_\eta^2 +3 (m_\eta r +1)] j_2 (qr).\label{dq}
\ee
In the zero momentum transfer limit\footnote{With however $m_\eta\neq 0$.
The limiting processes $q\rightarrow 0$ and $m_\eta\rightarrow 0$ do not
commute as we will see shortly.}, we have
\bq
A^j_{1\eta}(0) = \frac{g f}{2M}\left[(y_{\eta}^2 +2(y_{\eta}+1))\delta_{j\,k} -
y_{\eta}^2 \hat{q}_{j} \hat{q}_{k} \right] e^{-y_{\eta}} \la S^k\ra
\label{1eta}
\eq
where $y_\eta=m_\eta R$.

The boundary condition (\ref{bc3}) provides the relation between the 
quark and $\eta^\prime$ contributions. In the integrated form, (\ref{bc3})
is 
\bq
\int d\Sigma f x_3  \hat{r} \cdot\vec{\nabla} \eta = \int_{V_B} d^3 r
\frac{1}{2}  {\bar \Psi} \gamma_3 \gamma_5 \Psi
\eq
from which follows
\bq
\frac{g f}{M} = 3\, \frac{e^{y_\eta}}{y_\eta^2 + 2(y_\eta +1)}\, g_{A,quark}^0.
\label{ga2}
\eq
This is a Goldberger-Treiman-like formula relating the asymptotic pseudoscalar
coupling to the quark singlet axial charge.
From (\ref{1q}), (\ref{1eta}) and (\ref{ga2}), we obtain
\be
A_1^j=g_{A_1}^0 \la S^j\ra
\ee
with
\bq
g^0_{A_1}\label{A1}=\frac{gf}{3M}\frac{y_\eta^2 + 2(y_\eta +1)}{e^{y_\eta}}
=\frac 32 g_{A,quark}^0.
\label{ga3}
\eq
This is completely
analogous to the isovector axial charge $g_A^3$ coming from the bagged
quarks inside the bag plus the perturbative pion fields outside the bag.
Note that the singlet charge $g^0_{A_1}$ goes to zero when the bag is
shrunk to zero, implying that the coupling constant $g$ goes to
zero as $R\rightarrow 0$ as one can see from eq.(\ref{ga2}). 
This is in contrast to $g_A^3$ where
 the axial charge from the bag ``leaks'' into the hedgehog pion
outside the bag and hence even when the bag shrinks to zero, the isovector
axial charge remains more or less constant in agreement with the 
Cheshire Cat\cite{hosaka}.

An interesting check of our calculation of $\vec{A}_1$
can be made by looking at the $m_\eta\rightarrow 0$ limit. From (\ref{A1q})
and (\ref{1eta}), we find that our current satisfies
\bq
\hat{q}\cdot \vec{A}_1 (0)=\frac{gf}{M}(y_\eta +1) e^{-y_\eta}\la \hat{q}\cdot
\vec{S}\ra\label{conserve}
\eq
which corresponds to eq.(\ref{Dbag}). Now eq.(\ref{Dbag}) is an operator
equation so one can take the limit $m_\eta\rightarrow 0$ and expect the
right-hand side to vanish, obtaining $\hat{q}\cdot\vec{A}_1\rightarrow 0$.
Equation (\ref{conserve}) fails to satisfy this. The reason for this
failure is that the $q\rightarrow 0$ and $m_\eta\rightarrow 0$ limits do not
commute. To obtain the massless limit, one should take the $\eta^\prime$ mass
to go to zero first.

Before taking the zero-momentum limit,  the expression for $c(q)$ for
the $\eta$ field, (\ref{cq}), is
\bq
c(q) = \frac{fg}{3M}\left( \frac{m_\eta^2}{q^2}
\frac{e^{-y_\eta}(\cos({qr}) +\frac{m_\eta}{q}\sin({qr}))}{1 +
\frac{m_\eta^2}{q^2}}\right)
\eq
which vanishes in the  $m_\eta\rightarrow 0$ limit. On the other hand,
the $d (q)$, (\ref{dq}),  which before taking the zero-momentum limit,
 is of the form 
\be
d(q) &=& -\frac{fg}{2M}(\frac{e^{-y_\eta}(y_\eta^2 + 3y_\eta +3)}
{qR} j_1(qR)\nonumber\\
 &&- \frac{m_\eta^2}{q^2}e^{-y_\eta}(y_\eta +1)j_0(qR) +
 \frac{m_\eta^4}{q^4}
\frac{e^{-y_\eta}(\cos({qr}) +\frac{m_\eta}{q}\sin({qr}))}{1 +
\frac{m_\eta^2}{q^2}})
\ee
becomes in the $m_\eta\rightarrow 0$ limit
\bq
-\frac{fg}{2M}\frac{j_1(qR)}{qR}.
\eq
Adding the  quark current (\ref{A1q}) in the $q\rightarrow 0$ limit, we get
\bq
A_1^j(0) = \frac{fg}{M}(\delta^{jk} -\hat{q}^j\hat{q}^k) S_k
\eq
which satisfies the conservation relation. This shows that our formulas
are correct.

\subsubsection{The gluon current  $A_2^\mu$}
\indent\indent
The current $A_2^\mu$ involving the color gauge field is very intricate
because it is not possible in general
to write a gauge-invariant dimension-3 local operator
corresponding to the singlet channel. We will see however that it is
possible to obtain a consistent {\it axial charge} within the model. Here
we shall
calculate it to the lowest nontrivial order in the gauge coupling constant.
In this limit, the right-hand sides of the boundary conditions (\ref{E})
and (\ref{B}) can be dropped, reducing to the original MIT boundary 
conditions \cite{mit}. Furthermore the gauge field decouples from the other degrees of
freedom precisely because of the color anomaly condition that prevents
the color leakage, namely, the condition (\ref{bc2}). In its absence, this
decoupling could not take place in a consistent way.\footnote{To higher
order in the gauge coupling, the situation would be a lot more complicated.
A full Casimir calculation will be required to assure the consistency of
the procedure. This problem will be addressed in a future publication.}

We start with the divergence relation
\bq
\partial_\mu A^\mu_2 =
\frac{\alpha_s N_f}{2\pi}\sum_a \vec{E}^a \cdot \vec{B}^a \Theta_{V_B}.
\label{an2}
\eq
In the lowest-mode approximation, the color electric and magnetic fields are
given by
\bq
\vec{E}^a = g_s \frac{\lambda^a}{4\pi} \frac{\hat{r}}{r^2} \rho (r)
\label{ef}
\eq
\bq
\vec{B}^a = g_s \frac{\lambda^a}{4\pi}\left( \frac{\mu (r)}{r^3}(3 \hat{r}
\vec{\sigma} \cdot \hat{r} - \vec{\sigma}) + (\frac{\mu (R)}{R^3} + 2 M(r))
\vec{\sigma}\right)
\label{bf}
\eq
where $\rho$ is related to the quark scalar density $\rho^\prime$ as
\bq
\rho (r)=\int_\Gamma^r ds \rho^\prime (s)\label{density}\nonumber
\eq
and $\mu, M$ to the vector current density
\be
\mu (r) &=& \int_0^r ds \mu^\prime (s),\nonumber\\ 
M (r)&=& \int_r^R ds \frac{\mu^\prime (s)}{s^3}.\nonumber
\ee
The lower limit $\Gamma$ usually taken to be zero in the MIT bag model
will be fixed later on. It will turn out that what one takes for $\Gamma$
has a qualitatively different consequence on the Cheshire-Cat property
of the singlet axial current.
Substituting these fields into the RHS of eq.(\ref{an2}) leads to
\bq
\vec{q} \cdot \vec{A_2} = \frac{8 \alpha _s^2 N_f}{3\pi} \vec{\sigma} \cdot
\hat{q} \int_0^R dr \rho (r)\left(2\frac{\mu(r)}{r^3} + \frac{\mu
(R)}{R^3} + 2M(r)\right) j_1(qr)
\eq
where $\alpha_s = \frac{g^2_s}{4\pi}$ and we have used  $ \sum_{i\neq j}\sum_a
\lambda^a_i\lambda^a_j = - \frac{8}{3}$ for the baryons\footnote{Here we 
are making the usual assumption as in ref.\cite{jaffe} that the $i=j$ terms
in the color factor are to be excluded from the contribution on the ground
that most of them go into renormalizing the single-quark axial charge.
If one were to evaluate the color factor without excluding the 
diagonal terms using only the lowest mode, 
the anomaly term would vanish, which of course is incorrect. 
As emphasized in \cite{jaffe}, there may be residual finite
contribution with $i=j$ but no one knows how to compute this and so
we shall ignore it here. It may have to be carefully considered in a
full Casimir calculation yet to be worked out.}.

In order to calculate the axial charge, we  take the zero momentum limit and
obtain
\bq
\lim_{q \rightarrow 0} \vec{A_2}(\vec{q}) = \frac{8 \alpha_s^2 N_f}{9\pi}
\tilde{A}_2(R)\vec{S}
\label{A2}
\eq
where
\bq
\tilde{A}_2(R) = \int_0^R r dr \rho (r)\left(2M(r) + \frac{\mu(R)}{R^3} +
2\frac{\mu(r)}{r^3}\right)\equiv 2 \int^R_0 dr r\rho(r) \alpha(r).\label{A}
\eq
The quantity $\alpha(r)$ is defined for later purposes.
It is easy to convince oneself that (\ref{A2}) is gauge-invariant, i.e.,
it is $\propto \int_{V_B} d^3r \vec{r} \sum_a \vec{E}^a \cdot \vec{B}^a$
which is manifestly gauge-invariant. The result (\ref{A2}) was previously
obtained in \cite{hatsudazahed}.

\subsubsection{The Chern-Simons current and NRWZ counter term}
\indent\indent
The Chern-Simons current $K_\mu$ whose divergence is gauge-invariant
is not by itself gauge-invariant. The question that can be raised here
is: How is the gauge-invariant object (\ref{A2}) related to the Chern-Simons
current incorrectly used in refs.\cite{PVRB,PV}~?
To answer this question, we first take the $\lambda^a$ 
outside from the field operators
\be
G^a_\mu &=& \frac{g_s}{4\pi} \lambda^a {\cal G}_\mu,\nonumber\\
G^a_{\mu\nu}&=&  \frac{g_s}{4\pi} \lambda^a {\cal G}_{\mu\nu}.\label{curly}
\ee
This is convenient in abelianizing the theory. 

{}From the electric and magnetic fields in the cavity (see eqs. (\ref{ef}) and
(\ref{bf})) and using
\be
{\cal E}_i &=& -\partial_i {\cal G}_0,\\
{\cal B}_i &=& \varepsilon_{ i j k} \partial_j {\cal G}_k
\ee
we get, up to gauge transformations,

\be
{\cal G}_0(\vec{r}) &=&   \int_0^r
ds\frac{\rho (s)}{s^2},
\label{a0}\\
{\cal G}_i (\vec{r}) &=&  \left(\frac{\mu(r)}{r^3}
+ \frac{1}{2}\frac{\mu(R)}{R^3} +M(r)\right) (\vec{r} \wedge
\vec{\sigma})_i.
\label{ai}
\ee

The curly fields behave under gauge transformations as
\bq
{\cal G}^{\Lambda}_\mu = {\cal G}_\mu +\partial_\mu \Lambda.
\eq
Consider a static ${\cal G}_\mu$ and restrict ourselves to
time-independent field transformations. Then
\bq
\Lambda(\vec{r},t) = \Lambda_1 t + \Lambda_2 (\vec{r})
\eq
where $\Lambda_1$ is a constant so that
\bq
{\cal G}^{\Lambda}_0 = {\cal G}_0  + \Lambda_1
\eq
and $\Lambda_2 (\vec{r})$ is a time-independent function such that
\bq
{\cal G}^{\Lambda}_i = {\cal G}_i  + \partial_i \Lambda_2. 
\eq
For these fields the Chern-Simons current is given by

\bq
{\cal K}_i = -2 {\cal G}_0{\cal B}_i + 2 \varepsilon^{ i j k} {\cal G}_j
{\cal E}_k  + \it{O}(g_s^3). 
\label{ch}
\eq
At the surface of the bag 
\bq
\hat{r}\cdot {\cal \vec{K}} \sim  {\cal G}_0^{\Lambda} \hat{r}\cdot \vec{\cal B}
\eq
which is in general different from zero. We may choose the constant
$\Lambda_1$ so that ${\cal G}_0$ vanishes at the surface of the bag.
In general,
there may be a  finite contribution.  However this is no cause for
worry since the crucial point of our reasoning is that such a
contribution, if non-vanishing, will be 
canceled by the NRWZ surface counter term. 

The gauge dependence of the Chern-Simons current is given by
\bq
{\cal K}^\Lambda_i - {\cal K}_i \sim - \Lambda_1 \vec {\cal B}_i + (\vec {\partial} 
\Lambda_2 \wedge \vec {\cal E})_i
\label{kg}
\eq
where we have denoted by ${\cal E}$ and ${\cal B}$ the color 
electric and magnetic fields
with the $\lambda$ factor taken out as in eq.(\ref{curly}).
Since our fields are static ($ \vec{\partial} \wedge \vec{\cal{E}}= 0$), we may write the RHS of eq.(\ref{kg}) as an exact differential, i.e.,
\bq
\varepsilon_{ijk} \partial_j (\Lambda_2{\cal E}_k - \Lambda_1 {\cal G}_k).
\eq
This term when calculating the charge, i.e., integrating over the bag, will
be killed by the  NRWZ surface coupling. 
This shows that the Chern-Simons current cannot be injected into the
interior of the bag without properly imposing   
the NRWZ counter term, an error committed in refs.\cite{PVRB,PV}.
\subsubsection{The structure of the $\eta^\prime$}
\indent\indent
In our discussion on the boundary 
condition eq.(\ref{bc2}), 
we emphasized the role of the NRWZ mechanism in removing gauge-non-invariant
terms accumulating on the surface. An important point to note here is
that this mechanism
imposed no condition on the normal component of the Chern-Simons
current itself. It is just that the normal flux of the Chern-Simons
current was canceled by the surface counter term. 

Thus far we have assumed that the only relevant degrees of freedom are
the quarks and gluons inside the bag and the $\eta^\prime$ (and pions)
outside the bag. This is the minimal picture.
Now suppose that there are additional degrees of freedom 
(in addition
to (\ref{A1in}) and (\ref{A1out})) either outside or inside of the bag or 
both inside and outside. We shall assume for simplicity that there is one such
degree of freedom outside. The same result will be obtained for the
other cases except for possibly different physical interpretations.
Now from the divergence condition (\ref{ABJ}), 
the additional current must be gauge-invariant and divergenceless, i.e.,
\be
\delta A^\mu_{1\eta}=-\Delta^\mu \Theta_M
\ee
with 
\be
\partial_\mu\Delta^\mu=0.
\ee
A possible candidate for such a degree of freedom
could be a heavy quarkonium or a heavy gluonium.
The condition (\ref{Dbag}) would remain unchanged provided the boundary
condition (\ref{bc3}) is modified to
\bq
\frac 12\hat{n}\cdot({\bar{\Psi}}{\bf \gamma}\gamma_5\Psi) +
\hat{r}\cdot \vec{\Delta}
=f \hat{n}\cdot\d \eta. \label{bc4}
\eq

To see what the consequences of the boundary condition (\ref{bc4}) are, 
consider a
$\vec{\Delta}$ that can be written in terms of the harmonic function\footnote{
The argument given here is actually more general, applying as well to the
case where $\vec{\nabla}\times\vec{\Delta}\neq 0$.}
\bq
\vec{\Delta} = \vec{\nabla} \Phi 
\eq
with in the cavity
\bq
\Phi = \sum_{l,m} C_l r^lY_{l,m}.
\eq
The boundary condition, eq.(\ref{bc4}), gets a contribution from
$l=1$ and hence only the coefficient $C_1$ enters.
This modifies the  asymptotic normalization (\ref{ga2}) to
\bq
\frac{gf}{M} = \frac{e^{y_\eta}}{y_\eta^2 +2(y_\eta + 1)}
\left(\frac{3}{2} g^0_{A,quarks} + 4\pi R^4 c_1\right)
\label{an3}
\eq
where the normalization constant is chosen so that $\Phi(\vec{r}) =
c_1\vec{S}\cdot\vec{r}$. The new term contributing to  the singlet current in 
momentum space is given by
\bq
\Delta_i(q) =  4\pi c_1 R^2 S_i \frac{j_1(qr)}{q}. 
\label{d1}
\eq
This new current adds a contribution to $g_{A_1}^0$. 
Using eqs.(\ref{an3}) and (\ref{d1}) together, we obtain
\bq g_{A_1}^0 = \frac{gf}{3M}\frac{y_\eta^2 +2(y_\eta + 1)}{e^{y_\eta}}
=\frac{3}{2}( g^0_{A,quarks}+ \frac{4\pi}{3}R^4c_1).\label{ga1new} 
 \eq

The expression (\ref{ga1new}) has an interesting physical interpretation.
In terms of the $\eta\prime$
parameters, it is exactly the same as what we obtained before, i.e., 
eq.({\ref{ga3}). However this is not so in terms of the quarks and
the additional degree of freedom. 
One may interpret this as describing the quark-glueball
undressing  of the $\eta\prime$. It is not clear what this additional degree
of freedom could be: One could perhaps relate it to (a part 
of) the pseudoscalar field $G=\Tr G_{\mu\nu}\tilde{G}^{\mu\nu}$ (where 
$\tilde{G}_{\mu\nu}$ is dual to $G_{\mu\nu}$)
introduced by the authors of \cite{schechter}. Without knowing its content
or structure, one can however infer its role if one adopts the Cheshire-Cat
principle.  Equation (\ref{ga1new}) shows that $g^0_{A_1}$ will become
large in magnitude as the radius grows if $c_1$ is non-negligible
and this will violate the Cheshire Cat. Thus the Cheshire Cat will require
that $c_1\sim 0$. Since we do not know how to compute it within the model
anyway, we shall simply assume it to be zero.
An interesting possibility is that when the $\eta^\prime$
nucleon coupling is measured with accuracy, we will not only
determine $g_{A1}^0$ unambiguously but also 
learn more about this mysterious degree of freedom if it is not completely 
negligible.

\subsubsection{The two-component formula}
\indent\indent
The main result of this paper can be summarized in terms of 
the two component-formula for the singlet axial charge (with $c_1=0$),
\bq
g_A^0=g_{A_1}^0 +g_{A_2}^0 =\frac{3}{2} g^0_{A,quarks} + 
\frac{8 \alpha_s^2 N_f}{9\pi} A_2(R).\label{SV}
\eq
The first term is the ``matter'' contribution (\ref{A1})
and the second the gauge-field contribution (\ref{A2}). 
This is the chiral-bag version of Shore-Veneziano 
formula\cite{ward,schechter} relating
the singlet axial charge to a sum of an $\eta^\prime$ 
contribution and a glue-ball
contribution.

\section{Results}
\indent\indent
In this section, we shall make a numerical estimate of (\ref{A1}) and 
(\ref{A2}) in the approximation that is detailed above. In evaluating
(\ref{A1}), we shall take into account the Casimir effects due to the
hedgehog pions but ignore the effect of the $\eta^\prime$ field on
the quark spectrum. The interaction between the internal and external 
degrees of freedom occurs at the surface.  Our approximation consists 
of  neglecting in the expansion of the boundary condition  in powers of 
$\frac{1}{f}$ all $\eta$ dependence, i.e.
\bq
i \hat{r} \cdot \gamma \Psi =
e^{i\gamma_5 \vec{\tau}\cdot \hat{r}\frac{\varphi(\vec{r})}{f_\pi}}
e^{i\gamma_5 \frac{\eta}{f}}\Psi \sim 
e^{i\gamma_5 \vec{\tau}\cdot \hat{r}\frac{\varphi(\vec{r})}{f_\pi}}
\Psi
\label{qbc}
\eq 
This approximation is justified by the massiveness of the $\eta^\prime$
field in comparison to the Goldstone pion field that supports the
hedgehog configuration, $\varphi$. Within this approximation, 
we can simply take the numerical results from \cite{PVRB,PV} 
changing only the overall constants in front.

The same is true with the gluon contribution. To the lowest order
in $\alpha_s$, the equation of motion  for the gluon field is the
same as in the MIT bag model. This is easy to see, since the modified 
boundary conditions eqs.(\ref{E}) and (\ref{B}) become
\bq
\hat{r}_i G^{i \mu} = - \frac{\alpha_s N_F}{2 \pi}\frac{\eta}{f}\hat{r}_i 
{\tilde G}^{ i \mu} \sim 0.
\eq
The only difference from the MIT model is that
here the quark sources for the gluons are modified by the hedgehog pion
field in (\ref{qbc}). Again the results can be taken from 
\cite{PVRB,PV} modulo an 
overall numerical factor. 

In evaluating the anomaly contribution (\ref{A}), we face the same problem
with the monopole component of the ${{\vec E}^a}$ field as in 
\cite{PVRB,PV}. If we write 
\bq
{\bf  E}_i^a (r)=f(r) \hat{\bf r} \lambda_i^a
\eq
where the subscript $i$ labels the $i$th quark and $a$ the color,
the $f(r)$ satisfying the Maxwell equation is
\bq
f(r)=\frac{1}{4\pi r^2}\int^r_\Gamma ds \rho^\prime (s)
\equiv \frac{1}{4\pi r^2}\rho(r).\label{fr}
\eq
If one takes only the valence quark orbit -- which is our approximation,
then $\rho^\prime$ in the chiral bag takes the same form as in the
MIT model. However the quark orbit is basically modified by the
hedgehog boundary condition, so the result is of course not the same.
The well-known difficulty here is that the bag boundary condition for
the monopole component
\bq
\hat{\bf r }\cdot {{\bf E}_i^a}=0,\ \ \ \ {\rm at}\ \ r=R
\eq
is not satisfied for $\Gamma\neq R$. Thus as in \cite{PVRB,PV},
we shall consider both $\Gamma=0$ and $\Gamma=R$.

The existence of a solution which satisfies explicitly and
locally the boundary condition suggests an approach different from
the one in the original MIT calculation \cite{mit}, 
where the boundary condition of the electric field was imposed as an 
expectation value with respect to the physical hadron state. 
In \cite{mit}, the $E$ and $B$ field contributions 
to the spectrum were treated on a completely different footing. While in the 
former the contribution arising from the quark self-energies was included, 
thereby leading to the vanishing of the color electric energy, in the
latter they were not. This gave the color magnetic energy for the source of the 
nucleon-$\Delta$ splitting. We have performed a calculation for the energy with
the explicitly confined $E$ and $B$ field treated in a symmetric fashion
\cite{PV}. Although in this calculation the contribution of the color-electric 
energy was non-vanishing, it was found not to affect the nucleon-$\Delta$ mass 
splitting, and therefore could be absorbed into a small change of the 
unknown parameters, i.e., zero point energy, bag radius, bag pressure etc.
As we shall see shortly, the two ways of treating the confinement
with $\Gamma=R$ and $\Gamma=0$ give qualitatively different results
for the role of the anomaly. One could consider therefore that the
singlet axial charge offers a possibility of learning something about
confinement within the scheme of the chiral bag. At present, 
only in heavy quarkonia \cite{vz} does one have an additional
handle on these operators.

%
%
The numerical results for both cases are given in Table 1.

\begin{table}[tbh]
\caption[] {\small
The flavor-singlet axial charge of the
proton as a function of radius $R$ and the chiral angle $\theta$. The
column labeled $g_{A_1}^0$ corresponds to the total contribution from
the quarks inside the bag and $\eta^\prime$ outside the bag (eq.(\ref{A1})) and
 $g_{A_2}^0 (\Gamma=R)$  and ${g_{A_2}^0} (\Gamma=0)$ to the gluon contribution
eq.(\ref{A2}) evaluated with $\Gamma=R$ and
$\Gamma=0$ in (\ref{fr}), respectively.  The parameters are: $\alpha_s=2.2$, 
$m_\eta=958$ MeV and $f=93$ MeV. The row with $R=\infty$ corresponds to the
unrealistic (and extreme) case of
an MIT bag model with the same parameters for the same degrees of
freedom but containing {\it no pions}.}
\vskip .3cm
\begin{center}
\begin{tabular}{|c|c|c|c|c|c|c|}\hline
R(fm)&$\theta/\pi$&$g_{A_1}^0$&$g_{A_2}^0 (\Gamma=R)$&$g_{A_2}^0 (\Gamma=0)$&
$g_A^0 (\Gamma=R)$&$g_A^0 (\Gamma=0)$\\ \hline
0.0&-1.000&0.000&0.000&0.000&0.000&0.000\\
0.2&-0.742&0.033&-0.015&0.009&0.018&0.042\\
0.4&-0.531&0.164&-0.087&0.046&0.077&0.210\\
0.6&-0.383&0.321&-0.236&0.123&0.085&0.444\\
0.8&-0.277&0.494&-0.434&0.232&0.060&0.726\\
1.0&-0.194&0.675&-0.635&0.352&0.040&1.027\\ \hline
$\infty$&0.00&0.962& -1.277&0.804 &-0.297 &1.784 \\ \hline
\end{tabular}
\end{center}
\end{table}

\section{Discussion}
\indent\indent
The quantity we have computed here is relevant to two physical issues:
the so-called ``proton spin" issue and the Cheshire-Cat
phenomenon in the baryon structure. A more accurate result
awaits a full Casimir calculation which appears to be non-trivial.
However we believe that the qualitative feature of the given model
with the specified degrees of freedom will not be significantly
modified by the full Casimir effects going beyond the lowest order
in $\alpha_s$.

In the current understanding of the polarized structure functions of the
nucleon, the FSAC matrix element or the flavor-singlet axial charge of
the proton
is related to the polarized flavor-singlet structure function $\Delta\Sigma=\Delta u+\Delta d+\Delta s$ \cite{cheng,ellis}.
The presently available analyses give \cite{ellis,altarelli}
\be
\Delta\Sigma &=& 0.27\pm 0.04\pm \cdots \label{ellis}\\
&=& 0.10\pm 0.05({\rm exp})\pm^{0.17}_{0.11}({\rm th})
\label{altarelli}
\ee

Our predictions for $g_A^0$ -- which can be compared with $\Delta\Sigma$ --
differ drastically depending upon whether one takes
$\Gamma=0$ for which the color electric monopole field satisfies {\it only
globally} 
the boundary condition at the leading order (that is, as a matrix element 
between color-singlet states) as in the standard 
MIT bag-model phenomenology or $\Gamma=R$ which makes
the boundary condition satisfied locally.  The former configuration severely
breaks the Cheshire Cat with the bag radius $R$ constrained to
$0.5$ fm  or less (``little bag scenario") to describe the 
empirical values (\ref{ellis}) and
(\ref{altarelli}). This is analogous to what Dreiner, Ellis and Flores
\cite{DEF} obtained. In this scenario, there is no way that the Cheshire
Cat can be recovered in the singlet channel unless  a hitherto unknown
degree of freedom discussed above
which contributes the surface term $\hat{r}\cdot\vec{\Delta}$ in the
boundary condition (\ref{bc4}) intervenes massively with the right sign
to cancel the rest, a possibility which we find to be highly unlikely 
although not totally excluded.

On the other hand, the configuration with $\Gamma=R$ which
we favor
leads to a remarkably stable Cheshire Cat in consistency with other
non-anomalous processes where the Cheshire Cat is seen to hold 
within, say, 30\% \cite{PV,hosaka}. The resulting singlet axial charge
$g_A^0 <0.1$ is consistent with (\ref{altarelli}) though perhaps
 somewhat too low
compared with (\ref{ellis}). One cannot however take the near zero
value predicted here too literally since the value taken for $\alpha_s$
is perhaps too  large.  Moreover other  short-distance degrees of
freedom not taken into account in the model (such as the light-quark vector 
mesons and other massive mesons)
can make a non-negligible additional contribution\cite{schechter}. What is 
noteworthy is that there is a large cancellation between
the ``matter" (quark and $\eta^\prime$) contribution and the gauge field 
(gluon) contribution in agreement with the interpretation
anchored on $U_A (1)$ anomaly\cite{altarelli}.

As mentioned above -- and also noted in \cite{PVRB,PV}, the electric monopole
configuration with $\Gamma=R$ is non-zero at the origin and hence is 
ill-defined there. This feature does not affect, however, other phenomenology
as shown in \cite{PV}. We do not know yet
if this ambiguity can be avoided if other multipoles
and higher-order and Casimir effects are included in a consistent way.
This caveat notwithstanding, it seems reasonable to conclude from the result 
that if one accepts that the
singlet axial charge is  small {\it because of the cancellation}
in the two-component formula and if in addition
one demands that the Cheshire Cat hold
in the $U_A (1)$ channel {\it as in other non-anomalous sectors}, we are led
to (1) adopt the singular monopole configuration that  satisfies
the boundary condition {\it locally} and (2) to the possibility
that within the range of
the bag radius that we are considering, the $\eta^\prime$ is primarily
quarkish with $c_1\approx 0$.
This issue will be addressed further in a forthcoming publication which
will include Casimir effects.

\subsection*{Acknowledgments}
\indent\indent
We are grateful for helpful correspondence from Byung-Yoon Park.
This work was done while one of the authors (MR) was visiting the
Department of Theoretical Physics in the University of Valencia under
the auspices of ``IBERDROLA de Ciencia y Tecnologia." He is grateful 
for its support as well as for the hospitality of the members of the 
Theory Department.
\pagebreak


\begin{thebibliography}{99}
\bibitem{NRWZ1} H.B. Nielsen, M. Rho, A. Wirzba and I. Zahed,
\pl {\bf B269} (1991) 389 
\bibitem{NRWZ2} H.B. Nielsen, M. Rho, A. Wirzba and I. Zahed,
\pl {\bf B281} (1992) 345 
\bibitem{PVRB} B.-Y. Park,
V. Vento, M. Rho and G.E. Brown, \np {\bf A504} (1989) 829 
\bibitem{PV} B.-Y. Park and V. Vento, \np {\bf A513} (1990) 413
\bibitem{cheng} H.-Y. Cheng, ``The Status of the Proton Spin Problem," 
Lectures at the {\it Xth Spring School on Particles and Fields}, Taiwan,
ROC, March 20-22, 1996, hep-ph/9607254.
\bibitem{NRZ} See, e.g., {\it Chiral Nuclear Dynamics}\ by
M.A. Nowak, M. Rho and I. Zahed (World Scientific Pub. Singapore, 1996)
Chapter 8.
\bibitem{DEF} H. Dreiner, J. Ellis and R.A. Flores, \pl {\bf B221}
(1989) 167.
\bibitem{casimir} G.E. Brown, A.D. Jackson, M. Rho and V. Vento,
\pl {\bf B140} (1984) 285.
\bibitem{falomir} M.D. Francia, H. Falomir and E.M. Santangelo, \pl
{\bf B371} (1996) 285
\bibitem{hosaka} A. Hosaka and H. Toki, ``Chiral Bag Model for
the Nucleon,'' Phys. Repts. {\bf 277} (1996) 65.
\bibitem{mit}
T. De Grand, R.L. Jaffe, K. Johnson and J. Kiskis, Phys. Rev {\bf D12} (1975)
2060.
\bibitem{jaffe} R.L. Jaffe, \pl {\bf B365} (1996) 359.
\bibitem{hatsudazahed} T. Hatsuda and I. Zahed, \pl {\bf B221} (1989) 173.
\bibitem{schechter} J. Schechter, V. Soni, A. Subbaraman and H. Weigel,
\prl {\bf 65} (1990) 2955; Mod. Phys. Lett. {\bf AA5} (1990) 2543;
Mod. Phys. Lett. {\bf A7} (1992) 1
\bibitem{ward} G.M. Shore and G. Veneziano, \pl {\bf B244} (1990) 75
\bibitem{vz} M. Voloshin and V. Zakharov, Phys. Rev. Lett. {\bf 45} (1980)
688.
\bibitem{ellis} For summary, J. Ellis and M. Karliner, ``The Strange
Spin of the Nucleon," Lectures at the {\it Int. School of Nucleon Spin
Structure}, Erice, August 1995, hep-ph/9601280.
\bibitem{altarelli} G. Altarelli, R.D. Ball, S. Forte and G. Ridolfi,
hep-ph/9701289
\end{thebibliography}
\end{document}